\documentclass[11pt,twoside]{article}

\usepackage{asp2006}
\usepackage{epsf}
\usepackage{psfig}
\usepackage{lscape}
\usepackage{graphicx}

\bibpunct{(}{)}{;}{a}{}{,}

\begin{document}
\title{On the solar origin of the 220.7 signal}   
\author{A. Jim\'enez}   
\affil{Instituto de Astrof\'isica de Canarias, 38205, La Laguna, Tenerife, Spain}    
\author{R. A. Garc\'\i a}   
\affil{Laboratoire AIM, CEA/DSM-CNRS-Universit\'e Paris Diderot; CEA, IRFU, SAp, Centre de Saclay, F-91191, Gif-sur-Yvette, France}    
\begin{abstract} 
Gravity modes in the Sun have been long searched during the past decades. Using their asymptotic properties Garc\'\i a et al. (2007) found the signature of the dipole g modes analyzing an spectral window between 25 and 140 $\mu$Hz of velocity power spectrum obtained from the GOLF/SoHO instrument. Using this result it has been possible to check some properties of the structure of the solar interior (Garc\'\i a, Mathur \& Ballot 2008) as well as some indications on the dynamics of the core.
However, the individual detection of such modes remains evasive and they are needed to really improve our knowledge of the deepest layers in the Sun (Mathur et al. 2008). In this work we study the signal at 220.7 $\mu$Hz which is present in most of the helioseismic instruments during the last 10 years. This signal has been previously identified as part of a g-mode candidate in the GOLF data (Turck-Chi\`eze et al. 2004; Mathur et al. 2007) and in SPM/VIRGO (Garc\'\i a et al. 2008) with more than 90$\%$ confidence level. It could be labelled as the l=2 n=-3 g mode as it is in the region were this mode is expected.

We have checked the possibility that the 220.7 $\mu$Hz signal could have an instrumental origin without success by analysing all the available housekeeping data as well as the information on the roll, pith and yaw of the SoHO spacecraft. In consequence, we are confident that this signal has a solar origin.
\end{abstract}

\section{Data Sets and Analysis}
This study is based on data provided by the VIRGO/SoHO package (SPM, LOI, DIARAD \& PMO (Fr\"ohlich et al. 1995)) and complemented by measurements from GOLF/SoHO (Gabriel et al. 1995; calibrated into velocity following Ulrich el al. 2000 and Garc\'\i a et al. 2005) and ground based network GONG (Harvey et al. 1996). 4098-d time series (1996, April 11 to 2007, June 30) has been used. This time span has been divided into 66 series of 800 days with a 50 days shift every two consecutive ones. These 66 power spectra has been used to build the so-called time-evolution power diagrams. To remove the low-frequency trends in the time series, two different methods has been used: a 1 day running mean and a Backwards Difference Filter (properly correcting the power spectrum following Garc\'\i a \& Ballot 2008). Both methods yield to the same results. Finally,  a classical FFT an an Iterative Sine Wave Fitting has been used with similar results. 

\section{The signal at 220.7 $\mu$Hz}
A stable signal with time has been found at 220.7 $\mu$Hz in all the VIRGO package but with a better signal-to-noise (SNR) ratio in the SPMs (see Figure 1). This signal is clearly observed during all the 4098 days of the mission.  At around time series 60 the signal slightly shifts towards higher frequency.
 \begin{figure}[htb*]
    \centering
    \includegraphics[width=11cm]{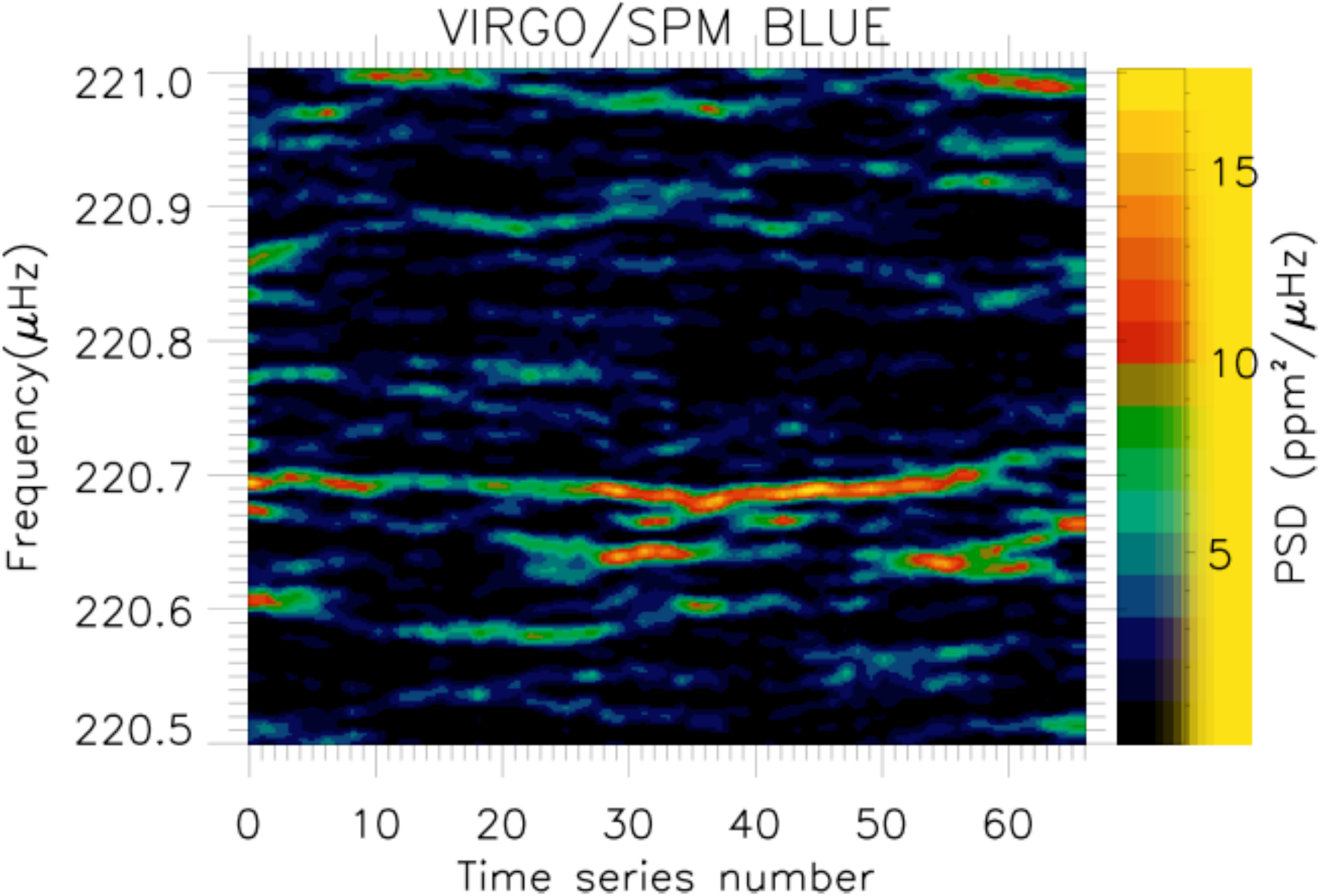}
    \caption{Time evolution power diagram of the SPM/VIRGO blue channel. A horizontal ridge with maximum power is visible at 220.7 $\mu$Hz}
\end{figure}

Averaging the signal on time we can produce a collapsed diagram in the frequency direction. Figure 2 showed such diagrams for three different instruments, GOLF and SPM aboard SoHO as well as the ground based network GONG. In all of them the highest peak in the region is the 220.7 $mu$Hz peak but with a poor SNR in the case of the ground-based data. However the presence of this peak in the GONG data minimise the possibility of a problem inside the SoHO spacecraft and it is more likely to have a solar origin.
 \begin{figure}[htbp*]
    \centering
    \includegraphics[width=11cm]{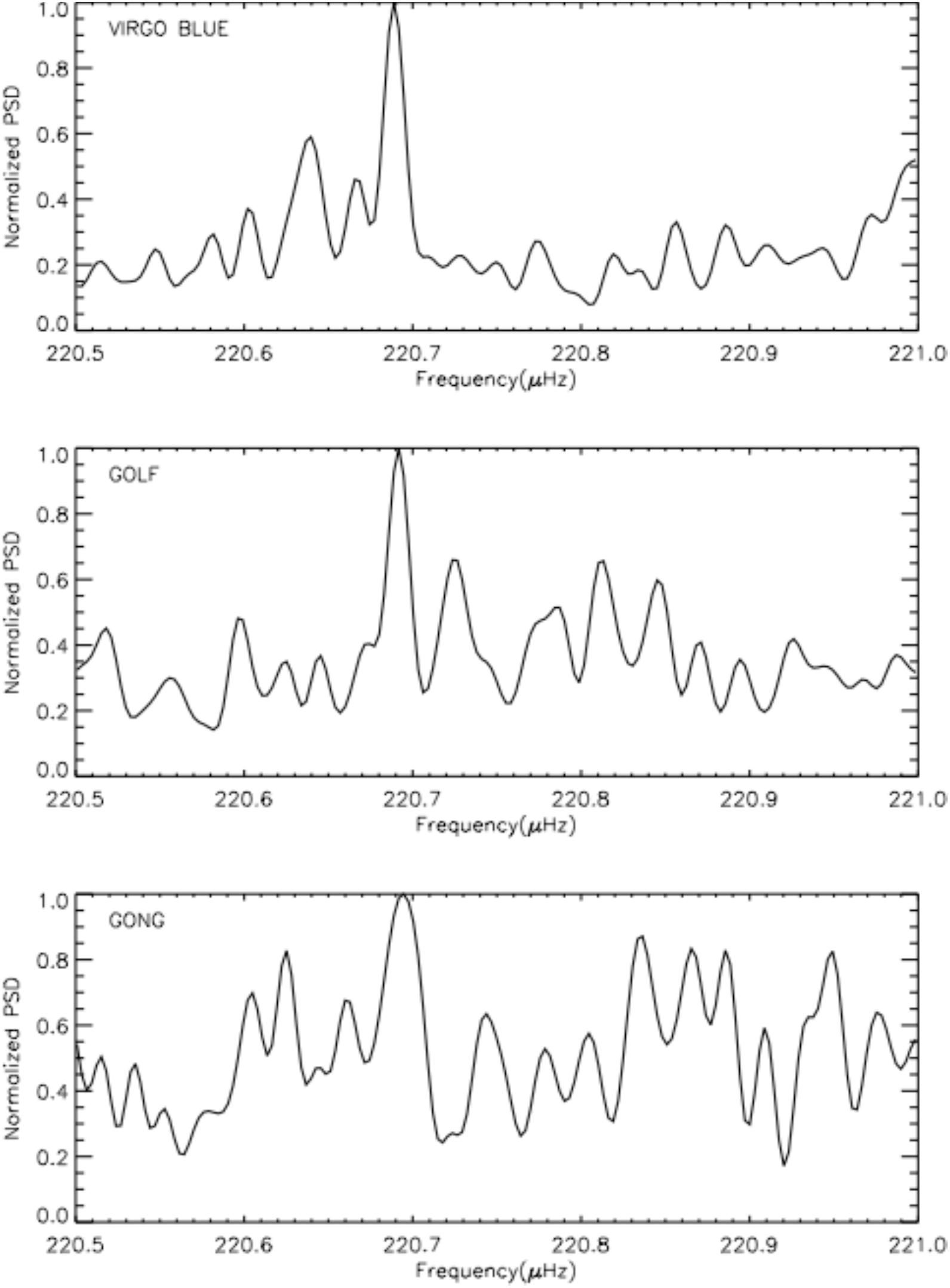}
    \caption{Collapsograms of the time evolution power diagrams computed using data from SPM, GOLF and GONG instruments. }
\end{figure}
\section{Looking into Housekeeping Data}
Once this interesting pattern have been detected in VIRGO the main question is to investigate its origin, that means, if it is a solar signal or a signal artificially produced by the instrumentation. Thus, we study all the possible non-solar origin of this signal which goes from the orbital and pointing corrections of the spacecraft to the housekeeping parameters of the VIRGO package. That means all the possible effects that could modulate the collected photometric signal. In particular, we have checked the temporal evolution of the following parameters yielding to the conclusion that neither of them could explain the observed pattern: Orbital corrections and Sun-Earth distance and velocity; SoHO spacecraft pointing: Pith, Yaw and even Roll; VIRGO Data Acquisition System (DAS) temperature variations; VIRGO Heatsink and DC/DC temperature variations; VIRGO/SPM sensor temperature corrections; VIRGO/SPM electronic temperature.

\section{Conclusions}
A signal at 220.7 is present during the last 12 years in all the instruments of the VIRGO/SoHO package
This signal has no instrumental origin. We have checked all the housekeeping parameters including the SoHO pointing, without success
A similar behaviour is found in the observations of the other instruments (i.e. GOLF), but with smaller S/N ratio 
The presence of this pattern in GONG points towards a solar origin.
Or this is the signature of an strange behaviour of the convection or it is a component of a g mode

\acknowledgements 
This work has been done inside the PHOEBUS collaboration funded by the International Space Science Institute in Bern, Switzerland. This work has also been partially founded by the Spanish grant PENAyA2007-62650 and the CNES/GOLF grant at the Sap-CEA/Sacaly. This work utilizes data obtained by the Global Oscillation Network Group (GONG) program, managed by the National Solar Observatory, which is operated by AURA, Inc. under a cooperative agreement with the National Science Foundation. The data were acquired by instruments operated by the Big Bear Solar Observatory, High Altitude Observatory, Learmonth Solar Observatory, Udaipur Solar Observatory, Instituto de Astrof\'\i sica de Canarias, and Cerro Tololo Interamerican Observatory. SOHO is an international cooperation between ESA and NASA.

\end{document}